\documentclass[12pt]{article}
\usepackage{amsmath,amsfonts,amssymb}

\begin{document}
\thispagestyle{empty}

\def\theequation{\arabic{section}.\arabic{equation}}
\def\a{\alpha}
\def\b{\beta}
\def\g{\gamma}
\def\d{\delta}
\def\dd{\rm d}
\def\e{\epsilon}
\def\ve{\varepsilon}
\def\z{\zeta}
\def\B{\mbox{\bf B}}

\newcommand{\h}{\hspace{0.5cm}}

\begin{titlepage}
%\vspace*{1.cm}
\renewcommand{\thefootnote}{\fnsymbol{footnote}}
\begin{center}
{\Large \bf Finite-Size Dyonic Giant Magnons in $TsT$-transformed
$AdS_5\times S^5$}
\end{center}
\vskip 1.2cm \centerline{\bf Changrim  Ahn }

\centerline{\sl Department of Physics} \centerline{\sl Ewha Womans
University} \centerline{\sl DaeHyun 11-1, Seoul 120-750, S. Korea}
\centerline{\tt ahn@ewha.ac.kr}

\vskip 0.6cm \centerline{\bf P.Bozhilov }

\centerline{\sl Institute for Nuclear Research and Nuclear Energy}
\centerline{\sl Bulgarian Academy of Sciences} \centerline{\sl  1784
Sofia, Bulgaria}

\centerline{\tt plbozhilov@gmail.com}

\vskip 20mm

\baselineskip 18pt

\begin{center}
{\bf Abstract}
\end{center}
\h We investigate dyonic giant magnons propagating on $\gamma$-deformed $AdS_5\times S^5$
by Neumann-Rosochatius reduction method with a twisted boundary condition. 
We compute finite-size effect of the dispersion relations of dyonic giant magnons which
generalizes the previously known case of the giant magnons with one angular momentum found by
Bykov and Frolov.

\end{titlepage}
%\end{quote}
%\vskip 1cm \centerline{\today}
\newpage
\baselineskip 18pt

%%%%%%%%%%%%%%%%%%%%%%%%%%%%%%%%%%%%%%%%%%%%%%%%%
\def\nn{\nonumber}
%%%%%%%%%%%%%%%%%%%%%%%%%%%%%%%%%%%%%%%%%%%%%%%%%
%%%%%%%%%%%%%%%%%%%%%%%%%%%%%%%%%%%%%%%%%%%%%%%%%%%%%
\def\tr{{\rm tr}\,}
\def\p{\partial}
\newcommand{\bea}{\begin{eqnarray}}
\newcommand{\eea}{\end{eqnarray}}
\newcommand{\bde}{{\bf e}}
\renewcommand{\thefootnote}{\fnsymbol{footnote}}
\newcommand{\be}{\begin{equation}}
\newcommand{\ee}{\end{equation}}
%\newcommand{\h}{\hspace{0.5cm}}
%%%%%%%%%%%%%%%%%%%%%%%%%%%%%%%%%%%%%%%%%%%%%%%%%%%%

\vskip 0cm

\renewcommand{\thefootnote}{\arabic{footnote}}
\setcounter{footnote}{0}

%\setcounter{equation}{0}
%\section{Introduction}

\setcounter{equation}{0}
%%%%%%%%%%%%%5%%%%%%%%%%%%%%%%%%%%%%%%%%%%%%%%%%%%%%%%%%%%%%%%%%%%%%%%%%%%%%%%
\section{Introduction}
%%%%%%%%%%%%%5%%%%%%%%%%%%%%%%%%%%%%%%%%%%%%%%%%%%%%%%%%%%%%%%%%%%%%%%%%%%%%%%
Investigations on AdS/CFT duality \cite{AdS/CFT,GKP98,EW98} for the
cases with reduced or without supersymmetry is of obvious interest
and importance. An interesting example of such correspondence
between gauge and string theory models with reduced supersymmetry is
provided by an exactly marginal deformation of $\mathcal{N} = 4$
super Yang-Mills theory \cite{LS95} and string theory on a
$\beta$-deformed $AdS_5\times S^5$ background suggested in
\cite{LM05}. When $\beta\equiv\gamma$ is real, the deformed
background can be obtained from $AdS_5\times S^5$ by the so-called
TsT transformation. It includes T-duality on one angle variable, a
shift of another isometry variable, then a second T-duality on the
first angle \cite{LM05,F05}. Taking into account that the
five-sphere has three isometric coordinates, one can consider
generalization of the above procedure, consisting of chain of three
TsT transformations. The result is a regular three-parameter
deformation of $AdS_5\times S^5$ string background, dual to a
non-supersymmetric deformation of $\mathcal{N} = 4$ super Yang-Mills
\cite{F05}, which is conformal in the planar limit to any order of
perturbation theory \cite{AKS06}. The action for this
$\gamma_i$-deformed $(i=1,2,3)$ gauge theory can be obtained from
the initial one after replacement of the usual product with
associative $*$-product \cite{LM05,F05,BR05}.

An essential property of the TsT transformation is that it preserves
the classical integrability of string theory on $AdS_5\times S^5$
\cite{F05}, which also implies that in the light-cone gauges of
\cite{AF04,FPZ06} the string dynamics on both backgrounds is
described by the same Hamiltonian density. The $\gamma$-dependence
enters only through the {\it twisted} boundary conditions and the
{\it level-matching} condition. The last one is modified since a
closed string in the deformed background corresponds to an open
string on $AdS_5\times S^5$ in general.

The finite-size correction to the giant magnon \cite{HM06}
energy-charge relation, in the $\gamma$-deformed background, has
been found in \cite{BF08}, by using conformal gauge and the string
sigma model reduced to $R_t\times S^3$. For the deformed case, this
is the smallest consistent reduction due to the {\it twisted}
boundary conditions. It turns out that even for the three-parameter
deformation, the reduced model depends only on one of them -
$\gamma_3$. As far as there are two isometry angles $\phi_1$,
$\phi_2$ on $S^3$, the solution can carry two non-vanishing angular
momenta $J_1$, $J_2$. Then, the giant magnon is an open string
solution with only one charge $J_1\ne 0$. The momentum $p$ of the
magnon excitation in the corresponding spin chain is identified with
the angular difference $\Delta\phi_1$ between the end-points of the
string, since in the light-cone gauge $t=\tau$, $p_{\phi_1}=1$, it
is equal to the worldsheet momentum $p_{ws}$ of a soliton
\cite{AFZ06}. The other angle satisfies the following {\it twisted}
boundary conditions \cite{BF08} \bea\nn
\Delta\phi_2=2\pi(n_2-\gamma_3 J_1),\eea where $n_2$ is an integer
winding number of the string in the second isometry direction of the
deformed sphere  $S_\gamma^3$.

%%%%% My addition
An interesting extension of this study is the dyonic giant magnon.
This state corresponds to bound states of the fundamental magnons and stable 
even in the deformed theory. 
Understanding its string theory analog in the strong coupling limit
can be helpful to extend the AdS/CFT duality to the deformed theories. 
%%%%%%%%%%%%%%%%%%%%%%%%%%%%%%

The paper is organized as follows. In Sect.2 we consider in brief
the $\gamma$-deformed giant magnon as described in the article by
Bykov and Frolov \cite{BF08}, and give their result about the
finite-size effect on the dispersion relation. In Sect.3 we
introduce the classical string action on $R_t\times S^3$, the
corresponding Neumann-Rosochatius (NR) integrable system and compute
the conserved quantities and angular differences for the case at
hand. In Sect.4 we provide our main result on the finite-size dyonic
giant magnon. We conclude the paper with some remarks in Sect.5.
Appendix A contains information about the elliptic integrals
appearing in the calculations, the $\epsilon$-expansions used and
the solutions for the parameters.

\setcounter{equation}{0}
%%%%%%%%%%%%%5%%%%%%%%%%%%%%%%%%%%%%%%%%%%%%%%%%%%%%%%%%%%%%%%%%%%%%%%%%%%%%%%
\section{The $\gamma$-deformed Giant Magnon}
%%%%%%%%%%%%%5%%%%%%%%%%%%%%%%%%%%%%%%%%%%%%%%%%%%%%%%%%%%%%%%%%%%%%%%%%%%%%%%
Our aim here is to briefly describe the explanations and the main
result derived in \cite{BF08}.

The bosonic part of the Green-Schwarz action for strings on the
$\gamma$-deformed $AdS_5\times S_\gamma^5$ \cite{AAF05} reduced to
$R_t\times S_\gamma^5$ can be written as (the common radius $R$ of
$AdS_5$ and $S_\gamma^5$ is set to 1) \bea\label{BGS}
S&=&-\frac{T}{2}\int d\tau
d\sigma\left\{\sqrt{-\gamma}\gamma^{ab}\left[-\p_a t\p_b t+\p_a
r_i\p_br_i+Gr_i^2\p_a\varphi_i\p_b\varphi_i \right.\right.\\
\nn &&+\left.\left.Gr_1^2r_2^2r_3^2
\left(\hat{\gamma}_i\p_a\varphi_i\right)
\left(\hat{\gamma}_j\p_b\varphi_j\right) \right]\right.\\ \nn
&&-2G\left.\epsilon^{ab}\left(\hat{\gamma}_3r_1^2r_2^2\p_a\varphi_1\p_b\varphi_2
+\hat{\gamma}_1r_2^2r_3^2\p_a\varphi_2\p_b\varphi_3
+\hat{\gamma}_2r_3^2r_1^2\p_a\varphi_3\p_b\varphi_1\right)\right\}
,\eea where $T$ is the string tension, $\gamma^{ab}$ is the
worldsheet metric, $\varphi_i$  are the three isometry angles of the
deformed $S_\gamma^5$, and \bea\label{roG} \sum_{i=1}^{3}r_i^2=1,\h
G^{-1}=1+\hat{\gamma}_3r_1^2r_2^2+\hat{\gamma}_1r_2^2r_3^2
+\hat{\gamma}_2r_1^2r_3^2.\eea The deformation parameters
$\hat{\gamma}_i$ are related to $\gamma_i$ which appear in the dual
gauge theory as follows \bea\nn \hat{\gamma}_i = 2\pi T \gamma_i =
\sqrt{\lambda} \gamma_i .\eea When $\hat{\gamma}_i=\hat{\gamma}$
this becomes the supersymmetric background of \cite{LM05}, and the
deformation parameter $\gamma$ enters the $\mathcal{N}=1$ SYM
superpotential in the following way \bea\nn W\propto
tr\left(e^{i\pi\gamma}\Phi_1\Phi_2\Phi_3-e^{-i\pi\gamma}\Phi_1\Phi_3\Phi_2\right).\eea

By using the TsT transformations which map the string theory on
$AdS_5\times S^5$ to the $\gamma_i$-deformed theory, one can relate
the angle variables $\phi_i$ on $S^5$ to the angles $\varphi_i$ of
the $\gamma_i$-deformed geometry \cite{F05}: \bea\label{r}
p_i=\pi_i,\h r_i^2\phi'_i=
r_i^2\left(\varphi'_i-2\pi\epsilon_{ijk}\gamma_j p_k\right),\h
i=1,2,3,\eea where $p_i$, $\pi_i$ are the momenta conjugated to
$\phi_i$, $\varphi_i$ respectively, and the summation is over $j,k$.
The equality $ p_i=\pi_i$ implies that the charges \bea\nn J_i=\int
d\sigma p_i \eea are invariant under the TsT transformation.

If none of the variables $r_i$ is vanishing on a given string
solution, from (\ref{r}) one gets \bea\nn \phi'_i=
\varphi'_i-2\pi\epsilon_{ijk}\gamma_j p_k.\eea Integrating the above
equations and taking into account that for a closed string in the
$\gamma$-deformed background \bea\nn
\Delta\varphi_i=\varphi_i(r)-\varphi_i(-r)=2\pi n_i,\h n_i\in
\mathbb{Z} ,\eea one finds the {\it twisted} boundary conditions for
the angles $\phi_i$ on the original $S^5$ space \bea\nn
\Delta\phi_i=\phi_i(r)-\phi_i(-r)=2\pi\left(n_i-\nu_i\right),\h
\nu_i=\varepsilon_{ijk}\gamma_jJ_k,\h J_i=\int_{-r}^{r} d\sigma p_i
.\eea It is obvious that if the {\it twists} $\nu_i$ are not
integer, then a closed string on the deformed  background is mapped
to an open string on $AdS_5\times S^5$.

The particular case considered in \cite{BF08} corresponds to
$J_2=J_3=0$, $\nu_1=0$, and as a result the angles $\phi_{1,2}$ of
the undeformed $S^3$ satisfy the following {\it twisted} boundary
conditions \bea\nn p=\Delta\phi_1=\phi_1(r)-\phi_1(-r),\h
\delta=\Delta\phi_2=\phi_2(r)-\phi_2(-r)=2\pi\left(n_2-\gamma_3J_1\right),\eea
where in fact $\delta$ plays the role of the deformation parameter.
By using the ansatz \bea\nn
&&\phi_1=\omega\tau+\frac{p}{2r}(\sigma-v\tau)+\phi(\sigma-v\tau),\\
\nn &&\phi_2=\nu\tau+\frac{\delta}{2r}(\sigma-v\tau)+\alpha(\sigma-v\tau),\\
\nn &&\chi=\chi(\sigma-v\tau),\eea where  $\phi$, $\alpha$ and
$\chi$ satisfy periodic boundary conditions, the authors of
\cite{BF08} found that the giant magnon string solution can be
completely determined from the equations \bea\nn &&\mathcal{E}\equiv
\frac{E_s}{\frac{\sqrt{\lambda}}{2\pi}}=2\int_{-r}^{0}d\sigma=2r,
\\ \nn &&\mathcal{J}_1=\frac{J_1}{\frac{\sqrt{\lambda}}{2\pi}}=
\frac{2}{1-v^2}\left(rv^2A_1+\omega\int_{\chi_{min}}^{\chi_{max}}d\chi\frac{1-\chi}{\vert\chi'\vert}\right),
\\ \label{ceq}
&&\mathcal{J}_2=\frac{J_1}{\frac{\sqrt{\lambda}}{2\pi}}\propto
rv^2A_2+\nu\int_{\chi_{min}}^{\chi_{max}}d\chi\frac{\chi}{\vert\chi'\vert}=0,
\\ \nn &&\frac{p}{2}+\frac{rv\omega}{1-v^2}=-\frac{vA_1}{1-v^2}
\int_{\chi_{min}}^{\chi_{max}}\frac{d\chi}{(1-\chi)\vert\chi'\vert},
\\ \nn &&\delta+\frac{rv\nu}{1-v^2}=-\frac{vA_2}{1-v^2}
\int_{\chi_{min}}^{\chi_{max}}\frac{d\chi}{\chi\vert\chi'\vert},
\eea where $A_1$ and $A_2$ are parameters related by $\omega A_1+\nu
A_2+1=0$, $\chi=1-r_1^2=r_2^2$, and \bea\nn \vert\chi'\vert=
\frac{2\sqrt{\omega^2-\nu^2}}{1-v^2}\sqrt{
(\chi_{max}-\chi)(\chi-\chi_{min})(\chi-\chi_{n})}, \\ \nn
0<\chi_{min}<\chi< \chi_{max}<1,\h \chi_{n}<0.\eea

The dispersion relation in the large $\mathcal{J}_1$ limit can be
found from (\ref{ceq}) as an expansion in \bea\nn
\exp\left(-\frac{\mathcal{J}_1}{\sin(p/2)}\right),\eea and up to the
leading order it is \cite{BF08} \bea\label{ecr1s} E-J_1=
\frac{\sqrt{\lambda}}{\pi}\sin(p/2)\left[1-\frac{4}{e^2}\sin^2(p/2)\cos(\Phi)
\exp\left(-\frac{\mathcal{J}_1}{\sin(p/2)}\right)\right],\eea where
\bea\nn \Phi=\frac{\delta}{2^{3/2}\cos^3(p/4)},\h
-\pi\le\delta\le\pi,\h -\pi\le p \le\pi .\eea In the limit $\Phi\to
0$ the formula (\ref{ecr1s}) reduces to the one obtained in
\cite{AFZ06}.

\setcounter{equation}{0}
%%%%%%%%%%%%%5%%%%%%%%%%%%%%%%%%%%%%%%%%%%%%%%%%%%%%%%%%%%%%%%%%%%%%%%%%%%%%%%
\section{Towards Finite-Size Dyonic Giant Magnon}
%%%%%%%%%%%%%5%%%%%%%%%%%%%%%%%%%%%%%%%%%%%%%%%%%%%%%%%%%%%%%%%%%%%%%%%%%%%%%%
As explained in the previous section, instead of considering strings
on the $\gamma$-deformed background $AdS_5\times S_\gamma^5$, we can
consider strings on the original $AdS_5\times S^5$ space, but with
{\it twisted} boundary conditions. Actually, here we are interested
in string configurations living in the $R_t\times S^3$ subspace,
which can be described by the NR integrable system \cite{KRT06}.

\subsection{Strings on $R_t\times S^3$ and the NR Integrable System}

We start with the Polyakov string action \bea &&S^P=
-\frac{T}{2}\int d^2\xi\sqrt{-\gamma}\gamma^{ab}G_{ab},\h G_{ab} =
g_{MN}\p_a X^M\p_bX^N,\\ \nn &&\p_a=\p/\p\xi^a,\h a,b = (0,1),
\h(\xi^0,\xi^1)=(\tau,\sigma),\h M,N = (0,1,\ldots,9),\eea and
choose {\it conformal gauge} $\gamma^{ab}=\eta^{ab}=diag(-1,1)$, in
which the Lagrangian and the Virasoro constraints take the form
\bea\label{l}
&&\mathcal{L}_s=\frac{T}{2}\left(G_{00}-G_{11}\right) \\
\label{00} && G_{00}+G_{11}=0,\qquad G_{01}=0.\eea

We embed the string in $R_t\times S^3$ subspace of $AdS_5\times S^5$
as follows \bea\nn Z_0=Re^{it(\tau,\sigma)},\h
W_j=Rr_j(\tau,\sigma)e^{i\phi_j(\tau,\sigma)},\h
\sum_{j=1}^{2}W_j\bar{W}_j=R^2,\eea where $R$ is the common radius
of $AdS_5$ and $S^5$, and $t$ is the $AdS$ time. For this embedding,
the metric induced on the string worldsheet is given by \bea\nn
G_{ab}=-\p_{(a}Z_0\p_{b)}\bar{Z}_0
+\sum_{j=1}^{2}\p_{(a}W_j\p_{b)}\bar{W}_j=R^2\left[-\p_at\p_bt +
\sum_{j=1}^{2}\left(\p_ar_j\p_br_j +
r_j^2\p_a\phi_j\p_b\phi_j\right)\right].\eea The corresponding
string Lagrangian becomes \bea\nn \mathcal{L}=\mathcal{L}_s +
\Lambda_s\left(\sum_{j=1}^{2}r_j^2-1\right),\eea where $\Lambda_s$
is a Lagrange multiplier. In the case at hand, the background metric
does not depend on $t$ and $\phi_j$. Therefore, the conserved
quantities are the string energy $E_s$ and two angular momenta
$J_j$, given by \bea\label{gcqs} E_s=-\int
d\sigma\frac{\p\mathcal{L}_s}{\p(\p_0 t)},\h J_j=\int
d\sigma\frac{\p\mathcal{L}_s}{\p(\p_0\phi_j)}.\eea

In order to reduce the string dynamics to the NR integrable system,
we use the ansatz \cite{KRT06} \bea\label{NRA}
&&t(\tau,\sigma)=\kappa\tau,\h r_j(\tau,\sigma)=r_j(\xi),\h
\phi_j(\tau,\sigma)=\omega_j\tau+f_j(\xi),\\ \nn
&&\xi=\alpha\sigma+\beta\tau,\h \kappa, \omega_j, \alpha,
\beta=constants.\eea It can be shown that after integrating once the
equations of motion for $f_a$, which gives \bea\label{fafi}
f'_a=\frac{1}{\alpha^2-\beta^2}
\left(\frac{C_a}{r_a^2}+\beta\omega_a\right),\h C_a=constants, \eea
one ends up with the following effective Lagrangian for the
coordinates $r_a$ (prime is used for $d/d\xi$) \bea\label{LNR}
L_{NR}=(\alpha^2-\beta^2)
\sum_{j=1}^{2}\left[r_j'^2-\frac{1}{(\alpha^2-\beta^2)^2}
\left(\frac{C_j^2}{r_j^2} + \alpha^2\omega_j^2r_j^2\right)\right]
+\Lambda_s\left(\sum_{j=1}^{2}r_j^2-1\right).\eea This is the
Lagrangian for the NR integrable system \cite{KRT06}.

The  Virasoro constraints (\ref{00}) give the conserved Hamiltonian
$H_{NR}$ and a relation between the embedding parameters and the
arbitrary constants $C_j$: \bea\label{HNR}
&&H_{NR}=(\alpha^2-\beta^2)
\sum_{j=1}^{2}\left[r_j'^2+\frac{1}{(\alpha^2-\beta^2)^2}
\left(\frac{C_j^2}{r_j^2} + \alpha^2\omega_j^2r_j^2\right)\right]
=\frac{\alpha^2+\beta^2}{\alpha^2-\beta^2}\kappa^2,
\\ \label{01R} &&\sum_{j=1}^{2}C_j\omega_j + \beta\kappa^2=0.\eea
On the ansatz (\ref{NRA}), $E_s$ and $J_j$ defined in (\ref{gcqs})
take the form \bea\label{cqs} E_s=
\frac{\sqrt{\lambda}}{2\pi}\frac{\kappa}{\alpha}\int d\xi,\h J_j=
\frac{\sqrt{\lambda}}{2\pi}\frac{1}{\alpha^2-\beta^2}\int d\xi
\left(\frac{\beta}{\alpha}C_j+\alpha\omega_j r_j^2\right),\eea where
we have used that the string tension and the 't Hooft coupling
constant $\lambda$ are related by
$TR^2=\frac{\sqrt{\lambda}}{2\pi}$.

\subsection{Conserved Quantities and Angular Differences}
If we introduce the variable \bea\nn \chi=1-r_1^2=r_2^2,\eea and use
(\ref{01R}), the first integral (\ref{HNR}) can be rewritten as
\bea\nn &&\chi'^{2}= \frac{4\omega_1^2(1-u^2)}{\alpha^2(1-v^2)^2}
\left\{-\chi^3+\frac{(1-w^2)+(1-v^2w^2)-u^2}{1-u^2}\chi^2\right.\\
\nn &&-\left. \frac{1-(1+v^2)w^2+v^2[(w^2-u^2j)^2-j^2]}{1-u^2}\chi-
\frac{v^2u^2j^2}{1-u^2}\right\}
\\ \label{eqchi} &&=\frac{4\omega_1^2(1-u^2)}{\alpha^2(1-v^2)^2}
(\chi_{max}-\chi)(\chi-\chi_{min})(\chi-\chi_{n}) ,\eea where
\bea\nn v=-\frac{\beta}{\alpha},\h u=\frac{\omega_2}{\omega_1},\h
w=\frac{\kappa}{\omega_1},\h j=-\frac{C_2}{\beta\omega_2}.\eea
Correspondingly, the conserved quantities (\ref{cqs}) transform to
\bea\nn &&\mathcal{E}=\frac{\kappa}{\alpha}\int_{-r}^{r}d\xi
=\frac{(1-v^2)w}{\sqrt{1-u^2}} \int_{\chi_{min}}^{\chi_{max}}
\frac{d\chi}{\sqrt{(\chi_{max}-\chi)(\chi-\chi_{min})(\chi-\chi_{n})}},
\\ \label{CQS} &&\mathcal{J}_1=\frac{1}{\sqrt{1-u^2}} \int_{\chi_{min}}^{\chi_{max}}
\frac{\left[1-v^2\left(w^2-u^2j\right)-\chi\right]d\chi}{\sqrt{(\chi_{max}-\chi)(\chi-\chi_{min})(\chi-\chi_{n})}},
\\ \nn &&\mathcal{J}_2=\frac{u}{\sqrt{1-u^2}} \int_{\chi_{min}}^{\chi_{max}}
\frac{\left(\chi-v^2j\right)d\chi}{\sqrt{(\chi_{max}-\chi)(\chi-\chi_{min})(\chi-\chi_{n})}}.\eea

Now, let us compute the angular differences \bea\nn
p=\Delta\phi_1=\phi_1(r)-\phi_1(-r),\h
\delta=\Delta\phi_2=\phi_2(r)-\phi_2(-r) =2\pi\left(n_2-\gamma_3
J_1\right).\eea \bea\label{dp}&&p=\int_{-r}^{r}d\xi
f'_1=\frac{\beta\omega_1}{\alpha^2(1-v^2)}\int_{-r}^{r}\left(1-\frac{w^2-u^2j}{r_1^2}\right)d\xi
\\ \nn &&=\frac{v}{\sqrt{1-u^2}}\int_{\chi_{min}}^{\chi_{max}}
\left(\frac{w^2-u^2j}{1-\chi}-1\right)\frac{d\chi}{\sqrt{(\chi_{max}-\chi)(\chi-\chi_{min})(\chi-\chi_{n})}},
\eea \bea\label{dd} &&\delta=\int_{-r}^{r}d\xi
f'_2=\frac{\beta\omega_2}{\alpha^2(1-v^2)}\int_{-r}^{r}\left(1-\frac{j}{r_2^2}\right)d\xi
\\ \nn &&=\frac{uv}{\sqrt{1-u^2}}\int_{\chi_{min}}^{\chi_{max}}
\left(\frac{j}{\chi}-1\right)\frac{d\chi}{\sqrt{(\chi_{max}-\chi)(\chi-\chi_{min})(\chi-\chi_{n})}}
.\eea

 The elliptic integrals
in (\ref{CQS}), (\ref{dp}) and (\ref{dd}) are given in Appendix A.

\setcounter{equation}{0}
%%%%%%%%%%%%%5%%%%%%%%%%%%%%%%%%%%%%%%%%%%%%%%%%%%%%%%%%%%%%%%%%%%%%%%%%%%%%%%
\section{Finite-Size Dyonic Giant Magnon}
%%%%%%%%%%%%%5%%%%%%%%%%%%%%%%%%%%%%%%%%%%%%%%%%%%%%%%%%%%%%%%%%%%%%%%%%%%%%%%
First of all, for correspondence with the notations in \cite{BF08},
we fix $\kappa=\alpha=1$, rename $\omega_1\to\omega$,
$\omega_2\to\nu$, introduce the parameters $A_1$, $A_2$, and the
functions $\phi(\xi)$, $\alpha(\xi)$ as follows \bea\nn
&&C_1=-vA_1,\h C_2=-vA_2, \\ \nn &&f_1(\xi)=\frac{p}{2r}\xi +
\phi(\xi),\h f_2(\xi)=\frac{\delta}{2r}\xi + \alpha(\xi).\eea Then,
from (\ref{CQS}), (\ref{dp}) and (\ref{dd}) one finds \bea\nn
&&\mathcal{E}=\frac{4\tilde{\kappa}}{\sqrt{(1-\chi_n)(1-\tilde{v}^2)}}\mathbf{K}(1-\epsilon),
\\ \nn
&&\mathcal{J}_1=\frac{4\tilde{\kappa}}{(1-v^2)\sqrt{(1-\chi_n)(1-\tilde{v}^2)}}
\left[\left(\omega (1-\chi_n)-\frac{v^2}{\omega}(1+\nu
A_2)\right)\mathbf{K}(1-\epsilon)\right. \\ \nn
&&-\left.\omega(1-\chi_n)
(1-\tilde{v}^2)\mathbf{E}(1-\epsilon)\right],
\\ \label{fexpr}
&&\mathcal{J}_2=\frac{4\tilde{\kappa}}{(1-v^2)\sqrt{(1-\chi_n)(1-\tilde{v}^2)}}
\left[\left(v^2 A_2+\nu\chi_n)\right)\mathbf{K}(1-\epsilon)\right. \\
\nn &&+\left.\nu(1-\chi_n)
(1-\tilde{v}^2)\mathbf{E}(1-\epsilon)\right],
\\ \nn &&p=\frac{4\tilde{\kappa}v}{(1-v^2)\sqrt{(1-\chi_n)(1-\tilde{v}^2)}}
\left[\frac{1+\nu A_2}{\omega(1-\chi_n)\tilde{v}^2}
\Pi\left(\frac{\tilde{v}^2-1}{\tilde{v}^2}(1-\epsilon)\vert
1-\epsilon\right)-\omega\mathbf{K}(1-\epsilon)\right],
\\ \nn &&\delta=-\frac{2\tilde{\kappa}v}{(1-v^2)\sqrt{(1-\chi_n)(1-\tilde{v}^2)}}
\left[\frac{A_2}{(1-\tilde{v}^2)\left(1+\chi_n\frac{\tilde{v}^2}{1-\tilde{v}^2}\right)}
\Pi\left(\frac{1-\chi_n}{1+\chi_n\frac{\tilde{v}^2}{1-\tilde{v}^2}}(1-\epsilon)\vert
1-\epsilon\right)\right. \\ \nn
&&+\left.\nu\mathbf{K}(1-\epsilon)\right], \\ \nn
&&\tilde{\kappa}=\frac{1-v^2}{2\sqrt{\omega^2-\nu^2}}.\eea In the
above equalities we introduced the new parameters \bea\nn
\tilde{v}^2=\frac{1-\chi_{max}}{1-\chi_{n}},\h
\epsilon=\frac{\chi_{min}-\chi_{n}}{\chi_{max}-\chi_{n}}\eea instead
of $\chi_{max}$ and $\chi_{min}$.

In order to obtain the finite-size correction to the energy-charge
relation, we have to consider the limit $\epsilon\to 0$ in
(\ref{fexpr}). The behavior of the complete elliptic integrals in
this limit is given in Appendix A. For the parameters in
(\ref{fexpr}), we make the following ansatz \bea\nn
&&v=v_0+v_1\epsilon+v_2\epsilon\log(\epsilon),\h
\tilde{v}=\tilde{v}_0+\tilde{v}_1\epsilon+\tilde{v}_2\epsilon\log(\epsilon),\h
\omega=1+\omega_1\epsilon,\\ \label{pex}
&&\nu=\nu_0+\nu_1\epsilon+\nu_2\epsilon\log(\epsilon),\h
A_2=A_{21}\epsilon,\h \chi_n=\chi_{n1}\epsilon.\eea We insert all
these expansions into (\ref{fexpr}) and impose the conditions:
\begin{enumerate}
\item{p - finite}
\item{$\mathcal{J}_2$ - finite}
\item{$\mathcal{E}-\mathcal{J}_1=\frac{2\sqrt{1-v_0^2-\nu_0^2}}{1-\nu_0^2}
- \frac{(1-v_0^2-\nu_0^2)^{3/2}}{2(1-\nu_0^2)}\cos(\Phi)\epsilon$}
\end{enumerate}
From the first two conditions, we obtain the relations
\bea\label{ctpj2}
p=\arcsin\left(\frac{2v_0\sqrt{1-v_0^2-\nu_0^2}}{1-\nu_0^2}\right),
\h \tilde{v}_0=\frac{v_0}{\sqrt{1-\nu_0^2}},\h
\mathcal{J}_2=\frac{2\nu_0\sqrt{1-v_0^2-\nu_0^2}}{1-\nu_0^2},\eea as
well as six more equations. The third condition gives another two
equations for the coefficients in (\ref{pex}). Thus, we have a
system of eight equations, from which we can find all remaining
coefficients in (\ref{pex}), except $A_{21}$. $A_{21}$ can be found
from the equation for $\delta$ to be \bea\nn
A_{21}=-\Lambda\frac{(1-v_0^2-\nu_0^2)^{3/2}}{v_0(1-\nu_0^2)}\sin(\Phi),\eea
where $\Lambda$ is constant with respect to $\Phi$. The equations
(\ref{ctpj2}) are solved by \bea\label{zms}
v_0=\frac{\sin(p)}{\sqrt{\mathcal{J}_2^2+4\sin^2(p/2)}},\h
\tilde{v}_0=\cos(p/2),\h
\nu_0=\frac{\mathcal{J}_2}{\sqrt{\mathcal{J}_2^2+4\sin^2(p/2)}}.\eea
Replacing (\ref{zms}) into the solutions for the other coefficients,
one obtains the expressions given in Appendix A.

To the leading order, the equation for $\mathcal{J}_1$ gives \bea\nn
\epsilon=16\exp\left[-\frac{2\left(\mathcal{J}_1 +
\sqrt{\mathcal{J}_2^2+4\sin^2(p/2)}\right)
\sqrt{\mathcal{J}_2^2+4\sin^2(p/2)}\sin^2(p/2)}{\mathcal{J}_2^2+4\sin^4(p/2)}
\right].\eea Accordingly, to the leading order again, the equation
for $\delta$ reads \bea\label{fd}
2\pi\left(n_2-\gamma_3\frac{\sqrt{\lambda}}{2\pi}\mathcal{J}_1\right)+
\frac{1}{2}\mathcal{J}_2\frac{\mathcal{J}_1 +
\sqrt{\mathcal{J}_2^2+4\sin^2(p/2)}}{\mathcal{J}_2^2
+4\sin^4(p/2)}\sin(p)=\Lambda\Phi .\eea Finally, the dispersion
relation, including the leading finite-size correction, takes the
form \bea\label{IEJ1} &&\mathcal{E}-\mathcal{J}_1 =
\sqrt{\mathcal{J}_2^2+4\sin^2(p/2)} - \frac{16 \sin^4(p/2)}
{\sqrt{\mathcal{J}_2^2+4\sin^2(p/2)}}\cos(\Phi)\\ \nn
&&\exp\left[-\frac{2\left(\mathcal{J}_1 +
\sqrt{\mathcal{J}_2^2+4\sin^2(p/2)}\right)
\sqrt{\mathcal{J}_2^2+4\sin^2(p/2)}\sin^2(p/2)}{\mathcal{J}_2^2+4\sin^4(p/2)}
\right].\eea For $\mathcal{J}_2=0$, (\ref{IEJ1}) reduces to the
result found in \cite{BF08}.

\setcounter{equation}{0}
\section{Concluding Remarks}
In this paper we considered giant magnons with two angular momenta,
or dyonic giant magnons, propagating on $\gamma$-deformed
$AdS_5\times S_\gamma^5$, obtained from $AdS_5\times S^5$ by means
of a chain of TsT transformations. In the framework of the approach
used in \cite{BF08}, instead of considering strings on the
$\gamma$-deformed background $AdS_5\times S_\gamma^5$, we considered
strings on the original $AdS_5\times S^5$ space, but with {\it
twisted} boundary conditions. Restricting ourselves to the
$R_t\times S^3$ subspace, we determined the leading finite-size
effect on the dispersion relation. The obtained dispersion relation
is a generalization of the previously known one for the giant
magnons with one angular momentum, found by Bykov and Frolov in
\cite{BF08}.

It would be interesting to reproduce the energy-charge relation
(\ref{IEJ1}) by using the L$\ddot{u}$scher's approach \cite{L86}. To
this end, we need a generalization of the L$\ddot{u}$scher's
formulas for the case of nontrivial {\it twisted} boundary
conditions.

\section*{Acknowledgements}
This work was supported in part by KRF-2007-313- C00150, WCU Grant
No. R32-2008-000-101300 (C. A.), by NSFB VU-F-201/06, DO 02-257 (P.
B.), and by the Brain Pool program from the Korean Federation of
Science and Technology (2007-1822-1-1).

\def\theequation{A.\arabic{equation}}
\setcounter{equation}{0}
\begin{appendix}

\section{Elliptic Integrals, $\epsilon$-Expansions and Solutions for the Parameters}

The elliptic integrals appearing in the main text are given by
\bea\nn &&\int_{\chi_{min}}^{\chi_{max}}
\frac{d\chi}{\sqrt{(\chi_{max}-\chi)(\chi-\chi_{min})(\chi-\chi_{n})}}
=\frac{2}{\sqrt{\chi_{max}-\chi_{n}}}\mathbf{K}(1-\epsilon),\\ \nn
&&\int_{\chi_{min}}^{\chi_{max}} \frac{\chi
d\chi}{\sqrt{(\chi_{max}-\chi)(\chi-\chi_{min})(\chi-\chi_{n})}}
\\ \nn &&
=\frac{2\chi_{n}}{\sqrt{\chi_{max}-\chi_{n}}}\mathbf{K}(1-\epsilon)+2\sqrt{\chi_{max}-\chi_{n}}
\mathbf{E}(1-\epsilon),
\\ \nn &&\int_{\chi_{min}}^{\chi_{max}}
\frac{d\chi}{\chi\sqrt{(\chi_{max}-\chi)(\chi-\chi_{min})(\chi-\chi_{n})}}
=\frac{2}{\chi_{max}\sqrt{\chi_{max}-\chi_{n}}}
\Pi\left(1-\frac{\chi_{min}}{\chi_{max}}\vert 1-\epsilon\right),
\\ \nn&&\int_{\chi_{min}}^{\chi_{max}}
\frac{d\chi}{\left(1-\chi\right)\sqrt{(\chi_{max}-\chi)(\chi-\chi_{min})(\chi-\chi_{n})}}
\\ \nn &&=\frac{2}{\left(1-\chi_{max}\right)\sqrt{\chi_{max}-\chi_{n}}}
\Pi\left(-\frac{\chi_{max}-\chi_{min}}{1-\chi_{max}}\vert
1-\epsilon\right),\eea where \bea\nn
\epsilon=\frac{\chi_{min}-\chi_{n}}{\chi_{max}-\chi_{n}}.\eea

We use the following expansions for the elliptic integrals \cite{w}
\bea\nn &&\mathbf{K}(1-\epsilon)=
-\frac{1}{2}\log(\epsilon)\left(1+\frac{\epsilon}{4}+O(\epsilon^{2})\right)+\log(4)-\frac{1}{4}\left(1-\log(4)\right)\epsilon+O(\epsilon^{2}),
\\ \nn
&&\mathbf{E}(1-\epsilon)= 1-\epsilon\left(\frac{1}{4}-\log(2)\right)
\left(1+O(\epsilon)\right)-\frac{\epsilon}{4}\log(\epsilon)\left(1+O(\epsilon)\right),
\\ \nn &&\Pi(n|1-\epsilon)=\frac{1}{2}\left(\frac{\log(\epsilon)}{n-1}\left(1-\frac{n+1}{4(n-1)}\epsilon
+O(\epsilon^{2})\right)\right.\\ \nn &&+
\frac{\sqrt{n}\log\left(\frac{1+\sqrt{n}}{1-\sqrt{n}}\right)-\log(16)}{n-1}-
\left.\frac{\sqrt{n}\log\left(\frac{1+\sqrt{n}}{1-\sqrt{n}}\right)-(n+1)\log(4)+1}{2(n-1)^{2}}\epsilon+O(\epsilon^{2})\right).\eea

By using the equality \cite{PBM-III} \bea\nn
\Pi(\nu|m)=\frac{q_{1}}{q}\Pi(\nu_1|m)-\frac{m}{q\sqrt{-\nu\nu_1}}\mathbf{K}(m),\eea
where \bea\nn &&q=\sqrt{(1-\nu)\left(1-\frac{m}{\nu}\right)},\h
q_1=\sqrt{(1-\nu_1)\left(1-\frac{m}{\nu_1}\right)},\\ \nn
&&\nu=\frac{\nu_1-m}{\nu_1-1},\h \nu_1<0,\h m<\nu<1 ,\eea and the
above expansion for $\Pi(n|1-\epsilon)$, one can find the following
expansion \bea\nn &&\Pi(1-\alpha \epsilon|1-\epsilon)= \frac{\arctan
\left(\sqrt{\frac{1}{\alpha}-1}\right)}{\sqrt{\frac{1}{\alpha}-1}\alpha\epsilon}\\
\nn &+& \frac{1}{4}\left(1-\frac{2\arctan
\left(\sqrt{\frac{1}{\alpha}-1}\right)}{\sqrt{\frac{1}{\alpha}-1}}-\log\left(\frac{\epsilon}{16}\right)\right)
\\
\nn &-& \frac{4\alpha^2\sqrt{\frac{1}{\alpha}-1}\arctan
\left(\sqrt{\frac{1}{\alpha}-1}\right)+(1-\alpha)\left(2(1+\alpha)+(1+2\alpha)\log\left(\frac{\epsilon}{16}\right)\right)}
{8(1-\alpha)}\epsilon+O(\epsilon^{2}),\eea where $0<\alpha<1$.

We use the following expansions for the parameters \bea\nn
&&v=v_0+v_1\epsilon+v_2\epsilon\log(\epsilon),\h
\tilde{v}=\tilde{v}_0+\tilde{v}_1\epsilon+\tilde{v}_2\epsilon\log(\epsilon),\h
\omega=1+\omega_1\epsilon,\\ \nn
&&\nu=\nu_0+\nu_1\epsilon+\nu_2\epsilon\log(\epsilon),\h
A_2=A_{21}\epsilon,\h \chi_n=\chi_{n1}\epsilon.\eea

The explicit solutions for the coefficients above are given by
\bea\nn &&v_0=\frac{\sin(p)}{\sqrt{\mathcal{J}_2^2+4\sin^2(p/2)}},\h
\tilde{v}_0=\cos(p/2),\h
\nu_0=\frac{\mathcal{J}_2}{\sqrt{\mathcal{J}_2^2+4\sin^2(p/2)}},\\
\nn &&v_1=\frac{1}{4\left(\mathcal{J}_2^2+4\sin^4(p/2)\right)}
\left\{\frac{1}{\left(\mathcal{J}_2^2+4\sin^2(p/2)\right)^{3/2}}
\left[\cos(\Phi)\sin(p)\sin^2(p/2)\right.\right. \\ \nn
&&\times\left.\left.
\left(\mathcal{J}_2^4(\log(256)-4)-16(\log(16)-1)\sin^4(p/2)+8\mathcal{J}_2^2\log(2)\sin^2(p)
\right.\right.\right. \\ \nn
&&+\left.\left.\left.4\sin^2(p/2)\left(\mathcal{J}_2^2(\log(16)-3)+(\log(16)-1)\sin^2(p)\right)\right)\right]
-\frac{1}{\mathcal{J}_2^2+4\sin^2(p/2)}\right. \\
\nn &&\times\left.
\left[2\Lambda\mathcal{J}_2\left(4\left(\mathcal{J}_2^2+4\sin^2(p/2)\right)(\log(4)-1)\sin^2(p/2)
+\sin^2(p)\left(\mathcal{J}_2^2(1-\log(16))
\right.\right.\right.\right. \\ \nn
&&+\sin^2(p)-\left.\left.\left.\left.\log(16)\cos(p)(1-\cos(p))\right)\right)\sin(\Phi)\right]\right\},
\eea
\bea\nn
&&v_2=-\frac{\sin^2(p/2)}{8\left(\mathcal{J}_2^2+4\sin^2(p/2)\right)^2\left(\mathcal{J}_2^2+4\sin^4(p/2)\right)}
\left\{\sqrt{\mathcal{J}_2^2+4\sin^2(p/2)}\left[4\mathcal{J}_2^4+6\mathcal{J}_2^2-10
\right.\right. \\ \nn
&&+\left.\left.\left(15-4\mathcal{J}_2^2\right)\cos(p)-2\left(3+\mathcal{J}_2^2\right)\cos(2p)
+\cos(3p)\right]\sin(p)\cos(\Phi)\right.
\\ \nn
&&+\left.2\Lambda\mathcal{J}_2\left[\left(4\mathcal{J}_2^4+17\mathcal{J}_2^2+34\right)\cos(p)
-4\left(2+\mathcal{J}_2^2\right)\cos(2p)\right.\right. \\ \nn
&&-\left.\left.\left(2+\mathcal{J}_2^2\right)\cos(3p)+\cos(4p)-12\mathcal{J}_2^2-25\right]\sin(\Phi)\right\},
\eea \bea\nn
&&\tilde{v}_1=\frac{\sin(p/2)}{8\sqrt{\mathcal{J}_2^2+4\sin^2(p/2)}\left(\mathcal{J}_2^2+4\sin^4(p/2)\right)}
\left\{\sqrt{\mathcal{J}_2^2+4\sin^2(p/2)}\right.\\ \nn
&&\times\left.\left[\left(4\left(\mathcal{J}_2^2-2\right)\log(2)\cos(\Phi)-\mathcal{J}_2^2-2\right)
\sin(p)+\left(1+\log(16)\cos(\Phi)\right)\right.\right.\\ \nn
&&\times\left.\left.\left(\sin^2(p)+\sin(2p)\right)\right]
+8\Lambda\mathcal{J}_2\left(4\cos(p)+\cos(2p)-2\mathcal{J}_2^2-5\right)\log(2)\sin^2(p/2)\sin(\Phi)\right\},
\eea
\bea\nn
&&\tilde{v}_2=\frac{\sin(p/2)}{16\sqrt{\mathcal{J}_2^2+4\sin^2(p/2)}\left(\mathcal{J}_2^2+4\sin^4(p/2)\right)}
\left[\sqrt{\mathcal{J}_2^2+4\sin^2(p/2)}\right.\\ \nn
&&\times\left.\left(3-2\mathcal{J}_2^2-4\cos(p)+\cos(2p)\right)\sin(p)\cos(\Phi)\right.\\
\nn
&&-4\left.\Lambda\mathcal{J}_2\left(4\cos(p)+\cos(2p)-2\mathcal{J}_2^2+-5\right)\sin^2(p/2)\sin(\Phi)\right],\eea
\bea\nn &&\omega_1=\frac{2\sin^4(p/2)}{\mathcal{J}_2^2+4\sin^4(p/2)}
\left(\sin^2(p/2)\cos(\Phi)+\frac{\Lambda\mathcal{J}_2\sin(p)\sin(\Phi)}{\sqrt{\mathcal{J}_2^2+4\sin^2(p/2)}}\right),
\eea
\bea\nn
&&\nu_1=\frac{\sin^2(p/2)}{2\left(\mathcal{J}_2^2+4\sin^4(p/2)\right)}
\left\{\frac{1}{\left(\mathcal{J}_2^2+4\sin^2(p/2)\right)^{3/2}}
\left[\mathcal{J}_2\cos(\Phi)\right.\right.
\\ \nn &&\times\left.\left.\left(9-2\cos(p)\left(6+\mathcal{J}_2^2-8\log(2)\right)-20\log(2)
-4\mathcal{J}_2^2\left(\log(4)-1\right)\right.\right.\right.
\\ \nn
&&+
\left.\left.\left.\cos(2p)\left(\log(16)+3\right)\right)\sin^2(p/2)\right]
+\frac{\Lambda\sin(p)\sin(\Phi)}{\mathcal{J}_2^2+4\sin^2(p/2)}
\left[6\log(4)+2\log(4)\cos(2p)\right.\right.\\ \nn
&&-\left.\left.\mathcal{J}_2^2\left(\log(256)-2\right)-2\cos(p)\left(\mathcal{J}_2^2+\log(256)\right)\right]\right\},
\eea
\bea\nn
&&\nu_2=\frac{1}{4\left(\mathcal{J}_2^2+4\sin^2(p/2)\right)^{3/2}\left(\mathcal{J}_2^2+4\sin^4(p/2)\right)}
\left\{2\mathcal{J}_2\sin^4(p/2)\cos(\Phi)\right. \\ \nn
&&\times\left.\left(5+2\mathcal{J}_2^2-4\cos(p)-\cos(2p)\right)
+\frac{4\Lambda\sin^3(p/2)\cos(p/2)\sin(\Phi)}{\sqrt{\mathcal{J}_2^2+4\sin^2(p/2)}}\right.
\\ \nn &&\times\left.\left[2\mathcal{J}_2^4+\mathcal{J}_2^2-10+15\cos(p)-\left(\mathcal{J}_2^2+6\right)
\cos(2p)+\cos(3p)\right]\right\}, \eea
\bea\nn
&&A_{21}=-\Lambda\sin^2(p/2)\tan(p/2)\sin(\Phi),\\ \nn
&&\chi_{n1}=-\sin^2(p/2)\sin^2(\Phi/2).\eea

Let us give some details about the derivation of $A_{21}$ and
$\chi_{n1}$, which are zero for the undeformed case. In our third
condition on p. 8 \bea\nn
\mathcal{E}-\mathcal{J}_1=\frac{2\sqrt{1-v_0^2-\nu_0^2}}{1-\nu_0^2}
- \frac{(1-v_0^2-\nu_0^2)^{3/2}}{2(1-\nu_0^2)}\cos(\Phi)\epsilon\eea
we introduced the angle $\Phi$ to describe in a simple way the
change of the finite-size correction to the dispersion relation due
to the $\gamma$-deformation. However, $\Phi$ is not an independent
new variable. Solving the equations for our parameters, we found
\bea\label{cpr}
\chi_{n1}=-\frac{1-v_0^2-\nu_0^2}{1-\nu_0^2}\sin^2(\Phi/2)
=-\sin^2(p/2)\sin^2(\Phi/2),\eea i.e. we use $\Phi$ instead of
$\chi_{n1}$.

There is alternative way to obtain the above relation between
$\chi_{n1}$ and $\Phi$. After expanding in $\epsilon$, the leading
order of the equation for the angle $\delta$ is given by \bea\nn
\delta&=&\frac{A_{21}v_0}{\chi_{n1}\sqrt{1-v_0^2-\nu_0^2}}
\sqrt{\frac{-\chi_{n1}}{1-\frac{v_0^2}{1-\nu_0^2}+\chi_{n1}}}
\arctan\sqrt{\frac{-\chi_{n1}}{1-\frac{v_0^2}{1-\nu_0^2}+\chi_{n1}}}
\\ \label{ltd} &&+\frac{\nu_0v_0}{2\sqrt{1-v_0^2-\nu_0^2}}\log\left(\frac{\epsilon}{16}\right)
.\eea Let us point out that the second term in (\ref{ltd}) is zero
for the one-spin case, since $\nu_0=0$ means $J_2=0$. If we
introduce the angle $\Phi$ as \bea\nn
\frac{\Phi}{2}=\arctan\sqrt{\frac{-\chi_{n1}}{1-\frac{v_0^2}{1-\nu_0^2}+\chi_{n1}}},\eea
this gives (\ref{cpr}), and the first term in (\ref{ltd}) takes the
form \bea\label{A21}
-\frac{A_{21}v_0(1-\nu_0^2)}{\left(1-v_0^2-\nu_0^2\right)^{3/2}}
\Phi\csc\Phi.\eea If we impose the natural condition (\ref{A21}) to
be an angle proportional to the angle $\Phi$, this gives \bea\nn
A_{21}=-\Lambda\frac{(1-v_0^2-\nu_0^2)^{3/2}}{v_0(1-\nu_0^2)}\sin(\Phi)
=-\Lambda\sin^2(p/2)\tan(p/2)\sin(\Phi),\eea where $\Lambda$ does
not depend on $\Phi$.

\end{appendix}

\end{document}